\documentclass[aps,groupedaddres,twocolumn]{revtex4}
\usepackage{graphicx}
\begin{document}
\title{Hydration and mobility of HO$^-$(aq)}
\author{D. Asthagiri}
\affiliation{Theoretical Division, Los Alamos National Laboratory, Los Alamos, NM 87545}
\author{Lawrence R. Pratt\footnote{Corresponding author: Email: lrp@lanl.gov; Phone: 505-667-8624; Fax: 505-665-3909}}
\affiliation{Theoretical Division, Los Alamos National Laboratory, Los Alamos, NM 87545}
\author{J. D. Kress}
\affiliation{Theoretical Division, Los Alamos National Laboratory, Los Alamos, NM 87545}
\author{Maria A. Gomez}
\affiliation{Department of Chemistry,  Mt.~Holyoke College, South Hadley, MA 01075}
\date{\today}
\begin{abstract}
The hydroxide anion plays an essential role in many chemical and
biochemical reactions. But a molecular-scale description of its
hydration state, and hence also its transport, in water is currently
controversial.  The statistical mechanical quasi-chemical theory of
solutions suggests that $\mathrm{HO\cdot[H_2O]_3{}^-}$ is the
predominant species in the aqueous phase under standard conditions. 
This   result is in close agreement with recent spectroscopic studies on
hydroxide water clusters, and with the available thermodynamic hydration
free energies.  In contrast, a recent  {\em ab initio\/} molecular
dynamics simulation has suggested that $\mathrm{HO\cdot[H_2O]_4{}^-}$ is
the only dominant aqueous  solution species.  We apply adiabatic {\em ab
initio\/} molecular dynamics simulations, and find good agreement with
both the quasi-chemical theoretical predictions and experimental
results.  The present results suggest a picture that is simpler, more
traditional, but with additional subtlety. These coordination structures
are labile but  the tri-coordinate species is the prominent case. This
conclusion is unaltered with changes in the electronic density
functional. No evidence is found for rate-determining activated 
inter-conversion of a $\mathrm{HO\cdot[H_2O]_4{}^-}$  trap structure to
$\mathrm{HO\cdot[H_2O]_3{}^-}$, mediating hydroxide transport.  The view
of HO$^-$ diffusion  as the hopping of a proton hole has substantial
validity, the rate depending largely on the {\em dynamic disorder} of
the water hydrogen-bond network.
\end{abstract}
\maketitle

\section{Introduction}

A pre-eminent challenge in liquid state physics is the understanding of
aqueous phase chemical transformations on a molecular scale. Water
undergoes limited autoprotolysis, which is enhanced in the presence of
highly charged metals, such as Be$^{2+}$
\cite{cecconi:ic02,lrp:becpl03}.  Understanding the hydration and
transport of the autoprotolysis products, H$^+$ and HO$^-$, presents
unique and interesting challenges for molecular-scale theories of
solutions and for simulations.  In this paper we focus on HO$^-$(aq).

Since H$^+$ and HO$^-$ constitute the underlying aqueous matrix, it is
not unreasonable to expect that their transport in water is different
from the transport of other aqueous ions. This anomalous diffusion of
the H$^+$(aq) and HO$^-$(aq) has received extensive scrutiny over the
years (for example, \cite{Bernal:JCP33,Eigen:Angew64,Stillinger:tc78}),
but recently {\em ab initio} molecular dynamics (AIMD) capabilities have
evolved to provide new information on the solution condition and
transport of these species. Over a similar period of time, the
statistical mechanical theory of liquids (especially water) has also
become usefully more sophisticated (for example, \cite{lrp:apc02}).
These two approaches can be complementary, but in typical practice they
remain imperfectly connected (but see
\cite{lrp:becpl03,lrp:jacs00,lrp:fpe01,lrp:jpca02,lrp:ionsjcp03,chandler:sc01}).

In an initial AIMD study \cite{Pnello:JCP95}, HO$^-$(aq) was observed to
be tetra-hydrated during the course of the $\sim$6~ps long simulation. 
This complex had a lifetime of about 2-3~ps. An approximately
square-planar configuration was noted for this
$\mathrm{HO\cdot[H_2O]_4{}^-}$ complex. That study hinted  that
transport occurred when $\mathrm{HO\cdot[H_2O]_4{}^-}$ converted to a
tri-hydrated ($\mathrm{HO\cdot[H_2O]_3{}^-}$) species which has
hydrogen-bonding arrangements similar to those in liquid water.
 
A recent AIMD study \cite{Pnello:Nature02} reinforced the notion of a
$\mathrm{HO\cdot[H_2O]_4{}^-}$ species exclusively dominating the
equilibrium population distribution at infinite dilution.  The proposed
mechanism for HO$^-$(aq) transport was:   first the stable (and hence
inactive) $\mathrm{HO\cdot[H_2O]_4{}^-}$ converts to the active
$\mathrm{HO\cdot[H_2O]_3{}^-}$ species; then the H-bond between the
anion and one of the ligating water molecules shortens, thus identifying
a transient $\mathrm{HO\cdot[H_2O]^-}$ species; a shared proton is
transferred along that shortened bond, and a
$\mathrm{HO\cdot[H_2O]_3{}^-}$ species is reconstituted with the
hydroxide identity now switched; this active
$\mathrm{HO\cdot[H_2O]_3{}^-}$ species reverts back to an inactive
$\mathrm{HO\cdot[H_2O]_4{}^-}$ species completing one transport event.
Presumably, as in \cite{Pnello:JCP95},  the lifetime of the
$\mathrm{HO\cdot[H_2O]_4{}^-}$ species was 2-3~ps, but statistical
characterization was sketchy.

Discussions of a transport mechanism for HO$^-$(aq) typically focus on
Agmon's \cite{agmon:jcp96,agmon:cpl00} extraction of an activation
energy for hydroxide transport from the temperature dependence of the
experimental mobilities.  Near room temperature that empirical parameter
is about 3~kcal/mol but increases by roughly a factor of two for
slightly lower temperatures.  As a mechanical barrier this value, about
5-6 $\mathrm{k_\mathrm{B}T}$, may be low enough to require some subtlety
of interpretation \cite{drozdov:jcp00}; the observed temperature
sensitivity of the activation energy,  and particularly its
\emph{increase} with decreasing temperature,  supports that possibility.
We note that a standard inclusion of a tunneling correction would be
expected to lead to a \emph{decrease} of activation energy with
decreasing temperature.

However, Ref.~\cite{Pnello:Nature02} framed the consideration of HO$^-$
transport in terms of classical transition state theory and extracted an
activation energy from the gas-phase study of Novoa~\emph{et~al.\/} 
\cite{novoa:jpca97}. Ref.~\cite{Pnello:Nature02} also considered the
importance of tunneling in lowering the barrier for proton transfer by
performing path integral calculations. Their combined value of
3.1~kcal/mol was close to Agmon's estimate
\cite{agmon:jcp96,agmon:cpl00}. That experimental  number does reflect
the influence of the solution medium,  and the Arrhenius plots are 
non-linear. Additionally, the earlier gas-phase studies 
\cite{novoa:jpca97} have shown that  outer shell disposition of the
fourth water molecule $\mathrm{HO\cdot[H_2O]_3{}^-\cdot H_2O}$ is lower
in energy than $\mathrm{HO\cdot[H_2O]_4{}^-}$, in agreement with
subsequent experimental results \cite{johnson:sc03}; and, further,  with
 a  barrier of 2.5~kcal/mol for conversion
$\mathrm{HO\cdot[H_2O]_3{}^-\cdot H_2O}$ to
$\mathrm{HO\cdot[H_2O]_4{}^-}$, nearly 1.3~kcal/mol {\em greater} than
the barrier for the reverse process.

A follow-up classical AIMD study \cite{Tuckerman:jpcb02},  which treated
a 1.5~M solution  of KOD (1 molecule of OD$^-$ and 1 atom of K$^+$ in a
32 water molecule system), estimated the diffusion coefficient of OD$^-$
to be 2.1~{\AA}$^2$/ps.   A tunneling correction for the classical
treatment would be expected to increase this rate. The  experimental
diffusion coefficient for the light water  OH$^-$ case is about
0.5~{\AA}$^2$/ps at 298~K \cite{Bernal:JCP33}.

The  incongruities in \cite{Pnello:JCP95,Pnello:Nature02} were noted
recently by \cite{Klein:jpca02} in AIMD simulations of deuterated NaOD
and KOD hydroxide solutions with concentrations ranging between 1.5~M to
15~M \cite{Klein:jpca02}. Interestingly, they observed that
$\mathrm{HO\cdot[H_2O]_3{}^-}$ was well-represented in the population
distribution, and commented that  their results \cite{Klein:jpca02}
differed somewhat from those obtained in the previous ab initio study
\cite{Pnello:JCP95,Pnello:Nature02} 
which reported $\mathrm{HO\cdot[H_2O]_4{}^-}$ to
be the only dominant solvation structure. The distribution of hydration
numbers was markedly influenced by the cations there 
\cite{Klein:jpca02}.  Those results do not permit a consistent
extrapolation of HO$^-$(aq) properties to infinite dilution, but
$\mathrm{HO\cdot[H_2O]_4{}^-}$ was just a prominent structure, not the
only one.

It is clear that earlier simulations have not resolved the most
primitive question: What is the coordination state of HO$^-$(aq) in
water without extrinsic complications? Speculations regarding the
transport mechanism are  somewhat premature in the absence of a clear
understanding of this coordination number question. Here we focus on
that primitive question first, apply AIMD methods,  and find that
$\mathrm{HO\cdot[H_2O]_3{}^-}$ is a probable coordination number.  We
then discuss  the agreement of the present simulation results with
inferences based on (a)  molecular theory \cite{lrp:HO02,lrp:hoqca03},
(b) spectroscopic  \cite{johnson:sc03} and thermo-chemical measurements
\cite{speller:jpc86} on hydroxide water clusters, (c) spectroscopic
studies\cite{zatsepina:zsk71,zundel:jcsFT73,librovich:cp79,librovich:rjpc82} of {\em aqueous\/} HO$^-$, and (d) dielectric dispersion
measurements of aqueous HO$^-$ \cite{sipos:jpcb99}. Discussion of the
transport can be then framed within the dynamical disorder framework
\cite{ratner:jcp83,ratner:prb85,zwanzig:pra85,zwanzig:acc90}.

\section{Recent Theoretical and Experimental Background}\label{theory}

Recent experimental and theoretical results have  addressed the issue of
 the  coordination number of the aquo hydroxide ion.  On the theoretical
 side, the statistical mechanical quasi-chemical theory of solution has
been applied to HO$^-$ (aq) \cite{lrp:HO02,lrp:hoqca03}.   This formally
exact approach,   with roots in the work of Guggenheim
\cite{Guggenheim:35,guggenheim:prsa38} and Bethe \cite{bethe:prsa35},
acquires approximations as  applied.  But for both hydration of ions in
water and standard packing problems, simple approximations have been
proven effective
\cite{lrp:becpl03,lrp:jacs00,lrp:fpe01,lrp:jpca02,lrp:ionsjcp03,lrp:Fejpca98, lrp:mulGjacs97,lrp:mp98,lrp:ES99,lrp:cp00,lrp:jpcb01,lrp:apc02,lrp:SCMF03}.  In the quasi-chemical approach, the region around the solute
is partitioned into inner and outer shell domains. The inner shell,
where chemical effects are important, is treated quantum mechanically.
The outer shell contributions can be assessed using classical
force-fields or dielectric continuum models \cite{lrp:ionsjcp03}. The
theory permits a variational check of the partition
\cite{lrp:fpe01,lrp:h2oAIMD03}.   Quasi-chemical studies
\cite{lrp:HO02,lrp:hoqca03} have firmly suggested that
$\mathrm{HO\cdot[H_2O]_3{}^-}$ is the most prominent solution species
\cite{lrp:HO02,lrp:hoqca03}.  Those results are insensitive to the
choice of density functional  or {\em ab initio\/}  M{\o}ller-Plesset 2$^\mathrm{nd}$ order perturbation technique.

Recent cluster spectroscopic studies \cite{johnson:sc03} have observed
shell closure with formation of $\mathrm{HO\cdot[H_2O]_3{}^-}$   in the
hydration of HO$^-$.   The fourth water molecule initiates an outer 
shell around this cluster.  This identification of a shell-closure is in
agreement with thermo-chemical measurements that  show the same effect
\cite{speller:jpc86}. Thus theoretical considerations and experiments on
hydroxide-water clusters concur on the significance of
$\mathrm{HO\cdot[H_2O]_3{}^-}$ as the nominal full inner shell
structure.

\section{Ab initio Molecular Dynamics Simulations}

The AIMD simulations were carried out with the VASP
\cite{kresse:prb93,kresse:prb96} simulation program wherein the
Kohn-Sham equations are solved by usual matrix methods. Thus this
corresponds to adiabatic dynamics, with tolerance set by the convergence
criterion for the electronic structure calculation. The system comprises
a hydroxide anion in a periodic cube of 32 water molecules. The  box
size was  9.8788~{\AA} consistent with the experimental partial specific
volume of the HO$^-$(aq) \cite{marcus}. This system was initially
thermalized by about 10 ps of classical molecular dynamics (using SPC/E \cite{spce}
potentials) with a temperature of 300~K using  velocity scaling. The
dominant HO$^-$ coordination number during this phase was $n=5$.

In the first AIMD simulation, RUN1, we used a generalized gradient
approximation, PW91 \cite{perdew:91,perdew:92},  to electron density
functional theory. Ultrasoft pseudopotentials (US-PP)
\cite{vanderbilt,kresse:jpcm94} for oxygen were used to describe the
core-valence interaction, and an ultrasoft pseudopotential was used for
the hydrogen atoms as well\cite{kresse:prb93,kresse:prb96}. The valence
orbitals were expanded in plane waves with a kinetic energy cutoff of
395.76~eV.  RUN1 used a 1~fs timestep. SCF convergence was accepted when
the energy difference between successive iterations fell below
$1\times10^{-4}$~eV. Initially 1.5~ps of AIMD simulations were
performed at 300~K using velocity scaling to maintain the stated
temperature. The coordination number at the end of this 1.5~ps
simulation was $n=4$. The system was then equilibrated for 3.2~ps in the
NVE ensemble. Then random velocities were reassigned to give a
temperature of 300~K, and statistics collected for another 8.2~ps in the
NVE ensemble.  The mean temperature was 332~K.  The relative energy
fluctuation, $\sqrt{\delta E^2}/|\bar{E}|$, was $8.4\cdot10^{-5}$.  The
drift in the relative  energy was about $8\cdot10^{-6}$~ps$^{-1}$. These
values appear quite reasonable \cite{allen}.

To evaluate the role of chosen density functionals and the timestep, we
took the terminal configuration from RUN1 and replaced all the hydrogen  atoms by deuterium,
thus simulating the classical statistical mechanics of aqueous DO$^-$ in
D$_2$O. The timestep was also cut in half to 0.5~ps. After a short NVE
run with the PW91 density functional and US-PP pseudopotentials, we
started RUN2 and RUN3. In both these runs the pseudopotential treatment
of atoms was replaced by the projector augmented-wave
\cite{blochl:prb94,kresse:prb99} (PAW) treatment, which is thought to
handle difficult cases involving large electronegativity differences
with ``exceptional precision" \cite{kresse:prb99}.  Further, in
molecular bonding problems, this method is about as accurate as local
basis (such as Gaussian orbital) methods \cite{weare:jpca99}. In RUN2
and RUN3, the SCF convergence was accepted when the energy difference
between successive iterations fell below $1\times10^{-6}$~eV. (For
comparison, this is an order of magnitude smaller than the ``tight"
convergence in the Gaussian \cite{gaussian} suite of programs.)

RUN2 employed the PBE functional \cite{perdew:prl96} which is similar to
the PW91.   RUN2  is not considered further here because it produced a
predominant coordination number of $n=3$, just as RUN1 did. RUN3
employed the revised PBE functional \cite{yang:prl98}. The system was
equilibrated for 5.9~ps and a further production run of 5.9~ps
was conducted.  The mean temperature was 313~K.  $\sqrt{\delta
E^2}/|\bar{E}|$, was $2.0\cdot10^{-5}$.  The drift in the
relative  energy was about $5\cdot10^{-6}$~ps$^{-1}$.

\begin{figure}[h]
\begin{center}
\includegraphics[width=1.5in]{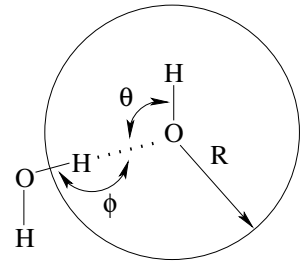}
\end{center}
\caption{$R$, is the radius of the observation volume centered on the
hydroxyl oxygen. $\theta$ and $\phi$ identify the angles that specify
the directionality of the H-bond to water.}\label{fg:selection}
\end{figure}

\begin{figure}[h!]
\begin{center}
\includegraphics[width=2.8in]{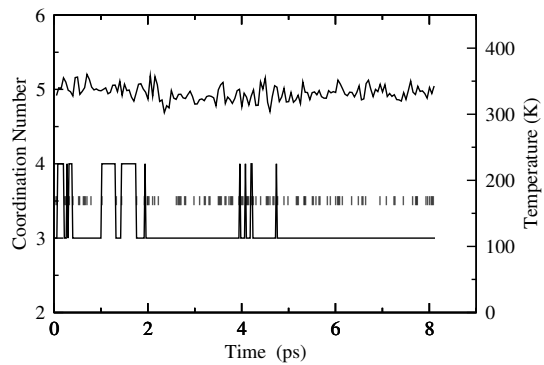}
\end{center}
\caption{RUN1 coordination number and temperature versus time.
$R=2.5$~{\AA}. The block averaged temperature is shown with the solid
line. The mean temperature is 332$\pm$22~K. The short vertical bars at
the $n=3.5$ level flag hydrogen exchange events, which also changes the
identity of the hydroxyl. Note that many hydrogen exchange events occur
without intercession of the n=4 configuration. }\label{fg:pw91c}
\end{figure}

\begin{figure}[h!]
\begin{center}
\includegraphics[width=2.8in]{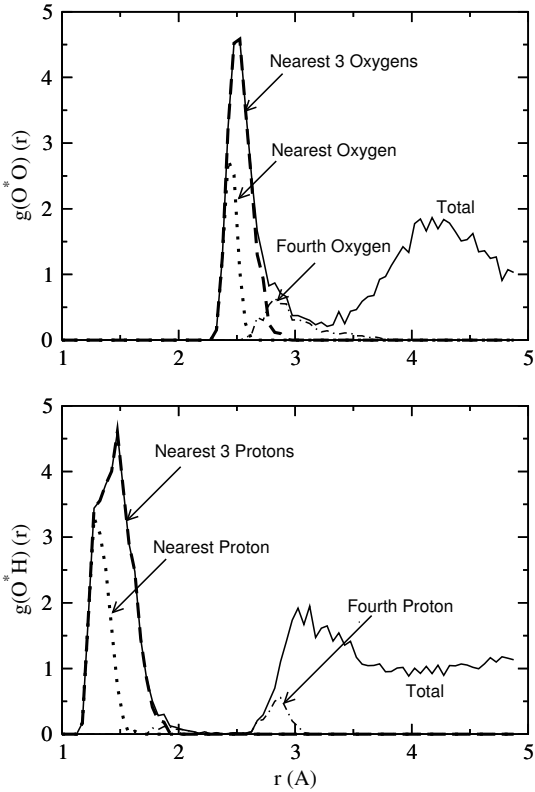}
\end{center}
\caption{Density distribution of water oxygen and proton around the
hydroxyl oxygen for RUN1 which utilized PW91. The distributions of the neighboring atoms
are also separated into contributions according to distance-order. The
hydrogen of the nominal HO chemical bond, otherwise the nearest H, isn't included in this distance-ordering.}\label{fg:pw91g}
\end{figure}

\begin{figure}[h!]
\begin{center}
\includegraphics[width=2.8in]{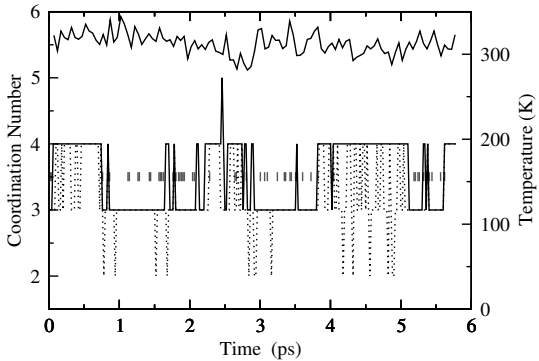}
\end{center}
\caption{RUN3  coordination number and temperature versus time. The mean
temperature is 313$\pm$21~K.  Rest as in FIG.~\ref{fg:pw91c}. The dashed
line applies to the selection criterion involving $R\leq2.5\, , \theta
\geq 80\, , \phi\geq 150$. Note that many hydrogen exchange events occur
without intercession of the n=4 configuration.}\label{fg:rpbec}
\end{figure}

\begin{figure}[h!]
\begin{center}
\includegraphics[width=2.8in]{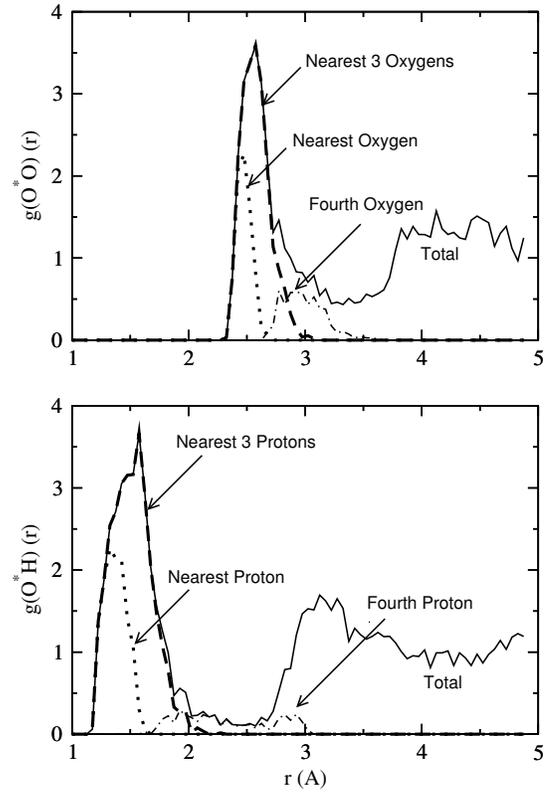}
\end{center}
\caption{Density distribution of water oxygen and proton around the
hydroxyl oxygen for RUN3 which utilized rPBE; otherwise as in Fig.~\ref{fg:pw91g}.}\label{fg:rpbeg}
\end{figure}

FIG.~\ref{fg:selection} introduces the geometric notation used in
analyzing the coordination of HO$^-$ (aq).   FIG.~\ref{fg:pw91c}  and
FIG.~\ref{fg:rpbec}  show the coordination number at each time for RUN1
and RUN3, and also the instantaneous temperature observed.  Radial
distribution functions for those two cases are shown  in
FIG.~\ref{fg:pw91g} and FIG.~\ref {fg:rpbeg}.  TABLE~\ref{tb:xs}
presents the average fractional coordination number populations for RUN1
and RUN3, and TABLE~\ref{TII} records averages of H-bonding angles using
the notation of FIG.~\ref{fg:selection}.

\begin{table}
\caption{Relative populations $\hat{x}_i= x_i/x_3$ for RUN1 and RUN3 and
different selection (FIG.~\ref{fg:selection}) criteria. The
quasi-chemical theory results are denoted by QC.}\label{tb:xs}
\begin{center}
\begin{tabular}{|l|ccccc|} 
\hline 
 Method  &    Criteria & $\hat{x}_1$ & $ \hat{x}_2$ & $\hat{x}_3$ & $ \hat{x}_4$ \\ \hline 
QC & --- &  0.03 & 0.26 & 1.0 &  0.0\\  \hline
 PW91/RUN1 & $R\leq2.5$ &--- & --- & 1.0 & 0.14 \\ 
  PW91/RUN1 & $R\leq2.5\, , \theta \geq 80\, , \phi\geq 150$ &--- &  0.03 & 1.0 & 0.08 \\ \hline
rPBE/RUN3 & $R\leq2.5$ &--- & --- & 1.0 & 1.02  \\
 rPBE/RUN3  &   $R\leq2.25$ &--- &  0.01 & 1.0 & 0.60  \\
 rPBE/RUN3   & $R\leq2.5\, , \theta \geq 80 $ & --- &  0.01 & 1.0 & 0.43 \\ 
 rPBE/RUN3   & $R\leq2.5\, , \theta \geq 80\, , \phi\geq 150$ & --- & 0.07 & 1.0 & 0.36 \\ \hline
\end{tabular}
\end{center}
\end{table}
\begin{table}
\caption{Mean angles $\theta$ and $\phi$, in degrees, defined in
Fig.~\ref{fg:selection},  for RUN1 and RUN3. The four values pertain to
the four nearest coordinating protons ordered in distance from the
anionic oxygen.  The values of  $\left\langle\theta_j\right\rangle$,
j=1,2, and 3 are consistent with classic tetrahedral geometry
(109.5$^\circ$).  These values are in good agreement with angles
(110$^\circ$) obtained from the optimized $\mathrm{HO\cdot[H_2O]_3{}^-}$
cluster.  $\left\langle\theta_4\right\rangle$ is different; $\theta_4$
is typically located closer to the equatorial plane, but with bigger
statistical dispersion.   $\left\langle\theta_4\right\rangle$ is also
different from the angle (116$^\circ$) obtained from the optimized
$\mathrm{HO\cdot[H_2O]_4{}^-}$ cluster.  The angles
$\left\langle\phi_j\right\rangle$ indicate that the coordinating OH bond
is not collinear with the O$^\ast$O vector, and this is consistent with
the cluster results. Note specifically, that the water oxygen atom determining
the angle $\phi_j$ doesn't corrrespond uniquely to a particular distance order for oxygen atoms;
this angle is defined by the distance-ordering of the hydrogen atoms, and the oxygen atoms
to which those hydrogens are directly bonded.}\label{tb:thetas}
\begin{center}
\begin{tabular}{|c|cccc|} 
 \hline
   & $\left\langle\theta_1\right\rangle$ & $\left\langle\theta_2\right\rangle$ & $\left\langle\theta_3\right\rangle$ & $\left\langle\theta_4\right\rangle$  \\ 
 \hline
 PW91/RUN1  & 107.9$\pm$9.2 & 108.1$\pm$10.4 & 107.0$\pm$11.8 &   97.4$\pm$32.8  \\ 
 rPBE/RUN3  & 109.1$\pm$9.9 & 108.0$\pm$11.9 & 103.4$\pm$13.1 &  91.8$\pm$24.5  \\ 
 \hline
    &  $\left\langle\phi_1\right\rangle$ & $\left\langle\phi_2\right\rangle$ & $\left\langle\phi_3\right\rangle$ & $\left\langle\phi_4\right\rangle$ \\ 
 \hline
  PW91/RUN1  &   169.2$\pm$9.9 & 168.1$\pm$6.3 &166.7$\pm$7.1 & 84.2$\pm$38.8\\ 
 rPBE/RUN3  &  169.8$\pm$5.4 & 168.0$\pm$6.0 & 165.0$\pm$7.9 &  124.9$\pm$45.8  \\ 
\hline
\end{tabular}
\end{center}
\label{TII}
\end{table}

\section{Discussion of AIMD Results}

For RUN1, it is clear that $n=3$ is the predominant state
(Figs.~\ref{fg:pw91c} and~\ref{fg:pw91g} and TABLE~\ref{tb:xs}). The
radial distribution functions shown in FIG.~\ref{fg:pw91g} (bottom
panel) establish that $R\leq 2.5$~{\AA}  is a reasonable, inclusive,
even permissive selection criterion; see also \cite{pimental}. 
TABLES~\ref{tb:xs} and~\ref{tb:thetas} provide guidance on whether many
of these $n=4$ configurations should be excluded as not hydrogen-bonded.
 Note that mediating $n=3$ to $n=4$  inter-conversions\ are not required
for exchange of the identity of the hydroxide oxygen.  RUN3 below
provided the same observation.

For RUN3, using the $R\leq2.5$~{\AA} criterion, we find about
equal populations of $n=3$ and $n=4$.  Tightening this criteria by 
0.25~{\AA} drops $n$=4 population by 40\% (TABLE~\ref{tb:xs}) relative to $n$=3. It is apparent
from TABLE~\ref{tb:thetas} that many of these $n=4$ states are  not 
square-planar. TABLE~\ref{tb:xs} shows that even a permissive
$\theta\geq80$ cutoff excludes many of those $n=4$ cases. The
configurations thereby excluded are on the `forward' side of the
hydroxide-water complex, above the plane defined by the hydroxyl oxygen
and  the shortest OH bond

The radial distributions (FIG.~\ref{fg:pw91g} and FIG.~\ref {fg:rpbeg})
decomposed according according to the distance-order of atoms
surrounding the hydroxide oxygen, denoted O$^\ast$, are similarly
interesting.  Though the statistical quality is meager, the conventional
O atom second shell begins at a distance roughly $\sqrt{2}$ times the radial
location of the first shell.    The  fourth-nearest
oxygen atom, qualitatively described, builds a shoulder on the outside of the principal maximum
of the O$^\ast$O radial distribution functions.  Note that the contributions from the nearest three protons and the nearest three oxygen are  concentrated, and that those protons are about 1\AA\ nearer the hydroxide oxygen.
In contrast, the contributions from the fourth-nearest proton and fourth-nearest oxygen are diffuse and overlapping; the contribution from the fourth-nearest proton is not always inside the contribution from the
fourth-nearest oxygen.   (Of course, those atoms needn't be directly bonded.)  These observations suggests again that  the fourth-nearest water molecule is not always participating in a conventional, specific H-bond but  is 
often non-specifically arranged. This description as a whole is consistent with the fact that the tri-coordinated species is a prominent species though the tetra-coordinated species is present to
some extent also.

The $\left\langle\theta_j\right\rangle$ results of TABLE~\ref{tb:thetas} document the interesting point that three nearest
coordinating protons are physically equivalent and approximately
disposed towards the corners of a tetrahedron. The fourth-nearest proton is
distributed broadly about the plane containing the hydroxyl oxygen and
perpendicular to the  OH chemical bond.  
 
Figs.~\ref{fg:pw91g} and~\ref{fg:rpbeg} show that the nearest
water-oxygen is always near 2.45$\pm0.1 (2\sigma)$~{\AA} of the hydroxyl
oxygen. This distance is close to the O-O separation in the calculated gas-phase
structure of $\mathrm{HO\cdot[H_2O]^-}$, 2.46{\AA}. We conclude that 
$\mathrm{HO\cdot[H_2O]^-}$ is a prominent sub-grouping in the
$\mathrm{HO\cdot[H_2O]_n{}^-}$ ($n=2,3,4$) species. The recent
theoretical and experimental work noted in Sec.~\ref{theory} suggests
that the structure $\mathrm{HO\cdot[H_2O]^-}$ does not provide as
natural a description  of the thermodynamic hydration free energy as
does the species $\mathrm{HO\cdot[H_2O]_3^-}$ that follow from a more
conservatively defined inner-shell. It seems likely that an inner-shell
definition designed \cite{lrp:cp00} to better isolate this
$\mathrm{HO\cdot[H_2O]^-}$ sub-grouping would be excessively
complicated.  This is one reason why $\mathrm{x_1}$ is not 
unambiguously separated by the results of TABLE~\ref{tb:xs}.

The identification of $\mathrm{HO\cdot[H_2O]^-}$ as a prominent
sub-grouping agrees with spectroscopic studies on concentrated hydroxide
solutions.  The IR and Raman spectra of concentrated hydroxide solutions
have been interpreted in terms of $\mathrm{HO\cdot[H_2O]^-}$ as a
principal structural possibility for those systems
\cite{zatsepina:zsk71,zundel:jcsFT73,librovich:cp79,librovich:rjpc82}.
But the present observations suggest that it is the  principal
participant in the proton conductivity.

This $\mathrm{HO\cdot[H_2O]^-}$  sub-grouping also concisely resolves
the high ``effective'' (not microscopic) hydration numbers extracted
from dielectric dispersion measurements \cite{sipos:jpcb99}.  A
`super'-grouping of hydrated $\mathrm{HO\cdot[H_2O]^-}$, one involving
several more water molecules, could well be relevant to the time scale
of the measurement, a possibility also suggested by Agmon
\cite{agmon:cpl00}. Then a dominating $\mathrm{HO\cdot[H_2O]_4{}^-}$
species \cite{Pnello:Nature02,Klein:jpca02,Tuckerman:jpcb02} is not a
conclusion necessary to the resolution of experimentally obtained
``effective'' hydration numbers.

The diffusion coefficient of HO$^-$ is calculated to be 3.1~{\AA}$^2$/ps
at 332~K using PW91 for HO$^-$, and 1.1~{\AA}$^2$/ps at 313~K using rPBE
for DO$^-$. For comparison, Zhu and Tuckerman's results for DO$^-$ gives
2.1~{\AA}$^2$/ps presumably at 300~K \cite{Tuckerman:jpcb02}.   The
experimental value, calculated from mobility data, is about
0.5~{\AA}$^2$/ps for HO$^-$  \cite{Bernal:JCP33}.  Perhaps the agreement
with experiment of this rPBE result is acceptable, but all these values
agree only roughly with experiments.  An important background point is
that evaluation of  diffusion coefficients  typically requires longer
simulation times, not readily accessible here by  AIMD simulation.

Rationalizations why the simulation rates are higher than observed 
experimental rates would be highly speculative.  A quantum mechanical
treatment of the water matrix, rather than a  classical  one, would
imply less order, and perhaps that would lead to slower rates.  Or it
might be that the electron density functionals used here lead to
excessive prominence of the $\mathrm{HO\cdot[H_2O]^-}$  sub-grouping,
and that leads to rates that are too high by comparison with
experiment.  The fact that the empirical activation energies
\emph{increase} with decreasing temperature remains an unexplained
point, apparently of qualitative significance.

The transport observed here involved the movement of a proton-hole
between two tri-hydrated species, which is also consistent with the
identification of the $\mathrm{HO\cdot[H_2O]^-}$ sub-grouping and the IR
spectra. Second-shell rearrangements do occur, but {\em
all-or-nothing\/} breaking and reforming of a hydrogen bond is not
necessary \cite{agmon:cpl00}.  There will certainly be rearrangements as
the hole settles into its new place. The hole-hopping proposal for 
HO$^-$ transport, as discussed by Bernal and Fowler \cite{Bernal:JCP33},
H{\"u}ckel \cite{Bernal:JCP33}, and later Stillinger
\cite{Stillinger:tc78}, has substantial validity.

We note that AIMD studies on pure liquid water \cite{lrp:h2oAIMD03}
under conventional thermodynamic conditions show that PW91 and PBE
predict more strongly structured liquid water compared to  experiment,
and rPBE softens the structure of liquid water simulated on that basis.
The excess chemical potential evaluated in that study
\cite{lrp:h2oAIMD03}, on the basis of AIMD results and the
quasi-chemical theory, indicates that PW91 binds liquid water too
strongly, whereas rPBE softens the binding. The excess chemical
potential of water at 314~K using the rPBE functional was
$-5.1$~kcal/mole (compared with $-6.1$~kcal/mole experimentally),
whereas PW91 gives $-12$~kcal/mole at 330~K.  Despite these differences,
the hydration structure of HO$^-$(aq) and  the qualitative transport
pattern are similar in these two cases. Therefore, the above comparisons
can be more optimistic for providing the correct direction to frame
discussions on HO$^-$ transport.

\section{Conclusion}

Three distinct lines of investigation, theory \cite{lrp:hoqca03},
experiments \cite{johnson:sc03,speller:jpc86}, and the present
simulations \cite{lrp:HO02}, converge on the common view that
$\mathrm{HO\cdot[H_2O]_3{}^-}$ is a prominent, likely even  dominating
coordination structure for HO$^-$ (aq); this is the most primitive  issue
underlying current speculations regarding HO$^-$ in aqueous solutions.  
The present simulation results suggest, in addition, that the 
coordination  number  distribution is labile, includes less specifically
structured $n$=4 possibilities, and that  $\mathrm{HO\cdot[H_2O]^-}$ is
a prominent sub-grouping within larger inner shell structures.  This
latter point is consistent   with interpretations of aqueous phase
spectroscopic
\cite{zatsepina:zsk71,zundel:jcsFT73,librovich:cp79,librovich:rjpc82}
results, and also with proposals of high-effective solvation numbers on
the basis of dielectric dispersion measurements \cite{sipos:jpcb99}.

\section{Acknowledgments}
We thank M.~L. Klein and M.~E. Tuckerman for their comments on this
work.  The work at Los Alamos was supported by the US Department of
Energy, contract W-7405-ENG-36, under the LDRD program at Los Alamos.
The work of MAG  was supported by a Camille and Henry Dreyfus Faculty
Start-up Grant Program for Undergraduate Institutions. LA-UR-02-7006.

\end{document}